\documentclass[showkeys, pra,aps,twocolumn,twoside,showpacs]{revtex4-2}
\usepackage[T1]{fontenc}
\usepackage{amsfonts}
\usepackage{amssymb}
\usepackage{mathrsfs}
\usepackage[intlimits]{amsmath}
\usepackage{bm, bbm}
\usepackage[dvips]{graphicx}

\usepackage[T1]{fontenc}
\usepackage[latin2]{inputenc}
\usepackage {hyperref}
\usepackage{amsmath}
\usepackage{latexsym}

\usepackage{color}
\usepackage{xcolor}
\usepackage{graphicx}
\usepackage[export]{adjustbox}
\usepackage{subcaption}
\usepackage{float}

\def\be{\begin{equation}}
\def\ee{\end{equation}}
\def\ben{\begin{eqnarray}}
\def\een{\end{eqnarray}}
\newcommand{\ket}[1]{| #1 \rangle}
\newcommand{\bra}[1]{\langle #1 |}

\hypersetup{
    colorlinks=true,
    linkcolor=blue,
    filecolor=magenta,      
    urlcolor=blue,
    pdftitle={Overleaf Example},
    pdfpagemode=FullScreen,
    }

\begin{document}
\title{Holevo bound and objectivity in the boson-spin model}

\author{Tae-Hun Lee}\email{Contact author: taehunee@cft.edu.pl}
\author{Jaros\l{}aw K. Korbicz}\email{Contact author: jkorbicz@cft.edu.pl}
\affiliation{\href{https://www.cft.edu.pl/}{Center for Theoretical Physics, Polish Academy of Sciences}, Aleja Lotnik\'ow 32/46, 02-668 Warsaw, Poland}
\date{\today}

\begin{abstract}
Emergence of objective, classical properties in quantum systems can be described in the modern language of quantum information theory. In this work, we present an example of such an analysis. We apply the quantum channel theory to a boson-spin model of open quantum systems and calculate, using recoilless approximation and the Floquet theory, the Holevo quantity, which bounds the capacity of the channel, broadcasting information about the central system into its environment. We analyze both the short-time regime, showing quadratic in time initial growth of the capacity, and the asymptotic regime. Complicated dependence on the model parameters, such as  temperature, tunneling energy for the environment, etc., is also analyzed, showing, e.g., regimes where the Holevo bound reaches its maximum.
\end{abstract}


\maketitle
\section{Introduction}
 
Most modern studies of how the objectivity of the macroscopic world appears during the  quantum-to-classical transition rely on the quantum Darwinism idea \cite{ZurekNature2009, ZurekPhysToday2014}. It is a more advanced form of a decoherence process and states that information about the system of interest, in order to become objective, has to be broadcast in multiple copies into the environment during the decoherence process.  This raises a natural question:  What is the capacity of such a broadcasting channel? This question has been studied in a series of works in spin-spin models, e.g., in models where a central spin interacts with a spin environment \cite{Zwolak_Riedel_Zurek:PRL2014, Wang:SciRep2015, Zwolak_Riedel_Zurek:SciRep2016, Zwolak_Zurek:PRA2017, Zwolak_Zurek:2019, Zwolak:2022ilt, Debarba:2024uhh}.  In this work, we complement those studies by considering a boson-spin model, where the central system is a harmonic oscillator. We will use the results obtained in our earlier studies of the model \cite{Lee:2024iso}. We briefly recall that the boson-spin model plays an important role as a simplified model representing interaction of electromagnetic field with two-level systems. It has been extensively studied with different combinations of couplings between the boson and the two-level systems and different approximation schemes. According to our focus, our boson-spin model can be categorized into the inhomogeneous Dicke model in quantum optics. Reviews of the previous studies of the model can be found, e.g., in Refs. \cite{Shirley:1965, Larson:2020inf, Garraway:2011, Keeling:2014,Larson:2007}.

Here, we approach the model from a different perspective than it has been traditionally studied, namely, that of information proliferation in the environment. Based on our earlier studies \cite{Lee:2024iso} of so-called spectrum broadcast structures (SBS) \cite{Korbicz:2014,Korbicz:2020tkf}, which are specific multipartite state structures responsible for a form of objectivity, we now study the process of objectification from the channel capacity point of view. The novel point compared to the earlier studies of the broadcasting capacity in this context \cite{Zwolak_Riedel_Zurek:PRL2014, Wang:SciRep2015, Zwolak_Riedel_Zurek:SciRep2016, Zwolak_Zurek:PRA2017, Zwolak_Zurek:2019, Zwolak:2022ilt, Debarba:2024uhh} is that now the central system is infinite-dimensional. Thus, we will investigate how well continuous-variable information is broadcast into the finite-dimensional spin environments. Since the infinite amount of continuous information cannot be stored into any range of finite degrees of freedom, encoding continuous information into finite environment should be understood as recorded information in finite bits with some finite resolution length \cite{Lee:2024iso}. As a result, the mutual information between the system with infinite degrees of freedom and one with finite degrees of freedom is understood as how many bits are used to encode continuous information with a given resolution.

In a broader context, these studies add to the exploration of the link between quantum information theory and quantum foundations. This important link has shed a new light on quantum foundations in the past years, e.g., highlighting the role of entanglement, just to name one example. Here, we examine the idea that objectification, necessary for quantum-to-classical transition, can be understood as a quantum-information process.

In particular, we are interested in the capacity of the effective quantum channels broadcasting system-related information into the environment during the open evolution and decoherence. Our main tool will be the well-known Holevo quantity $\chi(\rho)$, 
which bounds the mutual information  $I(X:Y)$, accessible in the environment \cite{Holevo:1973, Nielsen:2010}:
\begin{align}\label{mutual information}
 I(X:Y)\leq& S\left(\sum_Xp_X\rho_X\right)-\sum_Xp_XS(\rho_X)\\
 &\equiv \chi(\rho),\nonumber
\end{align}
where $S(\cdot)$ is the von Neumann entropy and $\rho_X$ are states encoding classical parameter $X$, distributed with some probability $p_X$.
We will calculate the continuous version of the Holevo bound  for a boson-spin model in the so-called recoilless limit, where the central oscillator is not much affected by the presence of the environment.
This is the opposite limit to the commonly studied Born approximation, see, e.g., Ref. \cite{Schlosshauer2007}, but is more suitable for objectivity studies where we are interested 
in how information flows from the system to the environment. The Holevo quantity can be calculated analytically, which as a by-product provides an interesting example of 
an exact, continuous-variable Holevo bound.  Since the analytical expressions are complicated, we will analyze both the short time and the asymptotic behavior, identifying how quickly the system-to-environment channel capacity grows and at what level it is stabilized, depending on the model parameters.

  \section{Dynamics of the system}
We start describing the system and  its approximate dynamics, based on Ref. \cite{Lee:2024iso}.
The Hamiltonian $H$ for the joint system of a harmonic oscillator of a mass $M$ and an angular frequency $\Omega$ and interacting qubits with a couplings $g_i$ and a self-energy $\Delta_i$ is given by 
 \begin{align}
H=H_{S}+\sum_iH^{(i)}_{E}+\sum_iH^{(i)}_{\text{int}},\label{H}
 \end{align}
 where
 \begin{align}\label{H_S,H_E,H_int}
    H_{S}&=\frac{\hat{P}^2}{2M}+\frac{1}{2}M\Omega^2 \hat{X}^2,\\
    H^{(i)}_E&=-\frac{\Delta_i}{2}\sigma^{(i)}_x,\\
    H^{(i)}_{\text{int}}&= g_i\hat{X}\otimes \sigma^{(i)}_z.
 \end{align}
Despite the apparent simplicity, the above dynamics is complicated and for the purpose of this work we will use a series of approximations, which nevertheless are enough to demonstrate the main features of the dynamics. First of all, since we are considering information transfer into the environment, it is suitable to assume that the state of the environment is allowed to change while the dynamics of the harmonic oscillator is not significantly influenced by interaction with the environment. This is called recoilless approximation and is the opposite of the broadly studied Born approximation. Negligence of the environmental recoil makes it possible to use the well-known Born-Oppenheimer approximation and the following ansatz can be used \cite{Tuziemski:2015}:  
\begin{equation}\label{final}
\ket{\Psi_{S:E}}=\int dX_0 \phi_0(X_0) e^{-i\hat H_St}|X_0\rangle U_{\text{eff}}[X(t;X_0)]\ket{\psi_0}.
\end{equation} 
Here $X(t;X_0)$ is an approximate trajectory followed by the central system and parametrized by the initial position $X_0$, 
$\phi_0(X_0) = \langle X_0|\phi_0\rangle$ is the wave function of the initial state of a central system, $\ket{\psi_0}$ is the  initial state of the environment, and the effective  evolution  $U_{\text{eff}}$
is governed by the corresponding effective Hamiltonian $H_{\text{eff}}$:
\begin{align}
     H_{\text{eff}}&=\sum_i\left[-\frac{\Delta_i}{2}\sigma^{(i)}_x+g_iX(t;X_0)\sigma^{(i)}_z\right].\label{Heff}
 \end{align}
Hence, the total unitary evolution operator is read by  
\begin{equation}\label{USE}
U_{S:E}(t)=\int dX_0 e^{-i H_St}\ket{X_0}\bra{X_0}\otimes U_{\text{eff}}[X(t;X_0)].
\end{equation}
Assuming a completely separable initial state,
\begin{align}\label{prod} 
\rho_{S:E}(0)=\rho_{S}(0)\otimes \bigotimes_i \rho_{E}^{(i)}(0),
\end{align}
the dynamics of the total density matrix $\rho_{S:E}(t)$ is written as
\begin{align}\label{rho_SE}
   \rho_{S:E}(t)&= \int dX_0dX'_0 \langle X_0|\rho_S(0)|X'_0\rangle e^{-iH_St}|X_0\rangle\langle X_0'| e^{iH_S t}\nonumber\\
& \otimes\bigotimes_{i=1}^N U_i(X_0,t)\rho^{(i)}_E(0) U^\dagger_i(X'_0,t),
\end{align}
where $U_i(X_0,t)$ corresponds to the effective Hamiltonian for the $i$th spin, $H_{i}$, in  $H_{\text{eff}}$ in \eqref{Heff}.
In particular, we will be interested in local states of environment fragments, called macrofractions. We thus trace out everything from the above state except for the chosen macrofraction (arbitrary at this moment):
\begin{align}\label{eq:rho_mac}
 & \rho_{\text{mac}}(t)=\text{Tr}_{S E \setminus \text{mac}}\rho_{S:E}(t)\nonumber\\
&=\int dX_0 p(X_0) \rho_{\text{mac}}(X_0),
\end{align}
where $p(X_0)\equiv \langle X_0|\rho_S(0)|X_0\rangle$ is the probability distribution of the initial position and
\be
 \rho_{\text{mac}}(X_0)\equiv \bigotimes_{i \in \text{mac}} U_i(X_0,t)\rho^{(i)}_E(0) U^\dagger_i(X_0,t)\label{mac}
\ee
is the conditional state corresponding to $X_0$. We will be calculating the Holevo quantity, associated with the ensemble \eqref{eq:rho_mac} and \eqref{mac}. 

\subsection{Initial state of the system}
Before we proceed, we need to specify the initial state of the system in order to obtain the approximate trajectories and the initial distribution $p(X_0)$; cf.\eqref{eq:rho_mac}. Especially, following the definition in Ref. \cite{Moller:1996}, we choose the displaced squeezed state as an appropriate way of controlling the initial probability distribution over the information transfer:
\begin{align}
\ket{\phi_0}=D(\alpha) S(\zeta)\ket{0} \equiv  |\alpha,\zeta,0\rangle,\ \alpha =|\alpha|e^{i\phi}, \zeta= re^{i\theta},
\end{align} 
where $D(\alpha)\equiv e^{\alpha a^\dagger-\alpha^*a}$ and $S(\zeta)\equiv e^{\frac{1}{2}(\zeta a^{\dagger2}-\zeta^* a^{2})}$ are the displacement and the squeezing operators respectively. Equivalently, a squeezed coherent state $S(\zeta)D(\tilde{\alpha})|0\rangle$ can be used with a substitution
\begin{align}
    \alpha \to \tilde{\alpha}=\alpha\cosh r-e^{i\theta}\alpha^*\sinh r.
\end{align}
In the absence of the environmental decoherence, the evolution of the above state would be simply given by $|\alpha,\zeta,t\rangle = e^{-iH_St}|\alpha,\zeta,0\rangle$,
leading to the well-known time-dependent Gaussian probability distribution of the position (see, e.g., Ref. \cite{Moller:1996})

\begin{align}\label{p(x0)}
    p_t(X)= |\langle X|\alpha,\zeta,t\rangle|^2&=\left[\frac{1}{\pi\eta(t)}\right]^{1/2} \exp\left\{-\frac{[X-q(t)]^2}{\eta(t)}\right\},
\end{align}
where
\begin{align}\label{eta(t),q(t)}
    \eta(t)&\equiv \frac{1}{M\Omega}[\cosh{2r}+\cos(2\Omega t + \theta)\sinh{2r}]>0,\\
    q(t)&\equiv \sqrt{\frac{2}{M\Omega}}|\alpha|\cos(\Omega t + \phi)
\end{align}
This motivates the following choice of the  trajectory in the Born-Oppenheimer ansatz \eqref{final}:
\begin{align}
    X(t;X_0)=X_0\cos(\Omega t+\phi), 
\end{align}
and defines the initial probability distribution through $p(X_0)=p_{t=0}(X_0)$.

\subsection{Floquet dynamics}
The effective Hamiltonian for the $i$th spin, $ H_{i}$, in  $H_{\text{eff}}$ in \eqref{Heff} is given by
\begin{align}
     H_{i}&=-\frac{\Delta_i}{2}\sigma^{(i)}_x+g_iX_0\cos(\Omega t+\phi)\sigma^{(i)}_z.\label{Hi}
 \end{align}
 The unitary evolution operator $U(X_0,t)$ on a single qubit can be calculated with the help of the Floquet theory and the high-frequency expansion as it was done, e.g., in Ref. \cite{Lee:2024iso}:
\begin{align}\label{U}
U(X_0,t)&=e^{-iK(t)}e^{-iH_{F}t}e^{iK(0)},
\end{align}
where 
\begin{align}
H_{F}t&=-\tilde{\Delta}(1-\xi^2)\tau\sigma_x+O(\xi^3),\label{He}\\
    K(t)&=\xi\sigma_z\sin\tau+O(\xi^2),\label{K}
\end{align}
and the following dimensionless parameter are introduced:
\begin{align}\label{dimensinless parameters}
  \tau\equiv\Omega t,~\xi\equiv g X_0/\Omega,~\tilde{\Delta}\equiv\Delta/2\Omega, 
\end{align}
together with a rescaled coupling strength:
\begin{align}\label{g_tilde}
    \tilde{g}\equiv\frac{g}{\Omega},
\end{align}
Operator \eqref{U} has the following Bloch representation:
\begin{align}\label{U_Bloch}
U(X_0,t)&=U_0\mathbb{I}+iU_1\sigma_x+iU_2\sigma_y+iU_3\sigma_z,
\end{align}
where
\begin{align}\label{U_i}\nonumber
 U_0&=\cos\{\xi[\sin(\tau+\phi)-\sin\phi]\}\cos[\tilde{\Delta}(1-\xi^2)\tau]\\\nonumber
U_1&=\cos\{\xi[\sin(\tau+\phi)+\sin\phi]\}\sin[\tilde{\Delta}(1-\xi^2)\tau]\\
U_2&=\sin\{\xi[\sin(\tau+\phi)+\sin\phi]\}\sin[\tilde{\Delta}(1-\xi^2)\tau]\\\nonumber
 U_3&=-\sin\{\xi[\sin(\tau+\phi)-\sin\phi]\}\cos[\tilde{\Delta}(1-\xi^2)\tau].
\end{align}

 We choose an initial state of the environment $ \rho_E(0)$ to be a thermal state for $H_E= -\Delta\sigma_x/2$ in \eqref{H}:
\begin{align}\label{thermal state}
    \rho_E(0)=\frac{1}{2}\left[\mathbb{I}+E(\beta)\sigma_x\right],
\end{align}
where $E(\beta)\equiv \tanh(\beta\Delta/2)$ with $\beta\equiv1/k_BT$.
Applying $U(X_0,t)$ from \eqref{U_Bloch} to \eqref{thermal state}, the final state $\rho_{X_0}(\tau)$ for a single-environmental qubit can be written in the Bloch representation as 
\begin{align}\label{rho_f}\nonumber
\rho_{X_0}(\tau)&=U(X_0,t)\rho_E(0) U^\dagger(X_0,t)\\
    &=\frac{1}{2}[\mathbb{I}+\vec{a}(X_0,\tau)\cdot\vec{\sigma}].
\end{align}
The explicit expressions $\vec{a}(X_0,\tau)$  were obtained in Ref. \cite{Lee:2024iso} and are shown in Appendix \ref{A:ai}.

 \section{Holevo quantity for a single environment}
 We consider the simplest case that the remaining spin after unobserved spin degrees of freedom traced out in the decoherence process, is an observed single spin. 
 
 According to Holevo's theorem, the quantum mutual information $I(X:Y)$ between a preparation system $X$ and a measurement system $Y$ is bounded by the Holevo quantity $\chi(\rho)$:
 \begin{align}\label{Holevo}
   I(X:Y)\leq \chi(\rho)\equiv S(\bar{\rho})-\bar{S},
\end{align}
where the average quantities $\bar{\rho}$ and $\bar{S}$ are defined by
\begin{align}\label{notation:rho_bar,S_bar}\nonumber
    \bar{\rho}&\equiv\int dX p(X)\rho_{X},\\
    \bar{S}&\equiv\int dX p(X) S(\rho_{X}).
\end{align}
Here, $\{\rho_{X}\}$ is a set of prepared states for random variables $X$ with the probability density distribution $\{p(X)\}$, and $S(\rho_X)$ is the entropy for a state $\rho_X$. Infinite dimensionality of $\rho_X$ could lead  $\chi(\rho)$ to be infinite unless physical constraints, e.g., the number of photons to be fixed, are applied \cite{Yuen:1993}. In our case $\rho_X$ has a finite dimension, so $\chi(\rho)$ is finite.

\subsection{Entropy of the average state}
Using \eqref{p(x0)} for $t=0$, the average density matrix $\bar{\rho}(\tau)$ for a qubit as introduced in \eqref{notation:rho_bar,S_bar} can be now calculated
\begin{align}\label{average rho}
\bar{\rho}(\tau)=\int dX_0 p(X_0)\rho_{X_0}(\tau).
\end{align}
$\bar{\rho}(\tau)$ is again written in the Bloch representation:
\begin{align}\label{rho_bar}\nonumber
     \bar{\rho}(\tau)&=\frac{1}{2}\int dX_0 p(X_0)[\mathbb{I}+\vec{a}(X_0,\tau)\cdot\vec{\sigma}]\\
     &\equiv\frac{1}{2}\left[\mathbb{I}+ E(\beta) \vec{\mu}(\tau)\cdot\vec{\sigma}\right],
\end{align}
where $\mu_i(\tau)$ are given by
\begin{align}\label{mui}\nonumber
    \mu_1(\tau)&\equiv\frac{1}{2}\text{Re}\{I[0,k_-,0]+I[0,k_+,0]+D_{12}(\tau)\}\\
   \mu_2(\tau)&\equiv\frac{1}{2}\text{Im}\{I[0,k_-,0]+I[0,k_+,0]+D_{12}(\tau)\}\\  
\mu_3(\tau)&\equiv\text{Re}[D_3(\tau)].\nonumber
\end{align}
The expressions for $I[\cdot,\cdot,\cdot]$ are given in \eqref{B:I(k)} and \eqref{B:I(dk)} and the decaying quantities $D_{12}(\tau)$ and $D_3(\tau)$ are defined as
\begin{align}\label{D12,D3}\nonumber
D_{12}(\tau)&\equiv \frac{1}{2}\{I[-k,k_-,-k_0]+I[k,k_-,k_0]\\
&-I[-k,k_+,-k_0]-I[k,k_+,k_0]\},\\
D_3(\tau)&\equiv \frac{1}{2}\{I[-k,\delta k/2,-k_0]-I[k,\delta k/2,k_0]\},\nonumber
\end{align}
with
\begin{align}\label{k}\nonumber
     &k_+=2\tilde{g}[\sin(\tau+\phi)+\sin\phi],\\\nonumber
    &k_-=2\tilde{g}[\sin(\tau+\phi)-\sin\phi],\\\nonumber
     &\delta k=k_+-k_-,\\
     &k=-2\tilde{g}^2\tau\tilde{\Delta},\\\nonumber
     &k_0=2\tau\tilde{\Delta}.
\end{align}
Note that the quantities $D_{12}(\tau)$ and $D_3(\tau)$ in \eqref{D12,D3} are a combination of $I[\pm k,\cdot,\cdot]$ with $k=-2\tilde{g}^2\tau\tilde{\Delta}\neq0$ and hence vanish as $t\to\infty$. The eigenvalues of $\bar{\rho}(\tau)$ in \eqref{rho_bar} are $\bar{\lambda}_{1,2}=[1\pm \mu(\tau) E(\beta)]/2$, $\mu(\tau)\equiv |\vec{\mu}(\tau)|$. 
Thus, the entropy $S(\bar{\rho})$ is expressed as
\begin{align}\label{S rho bar}\nonumber
    S(\bar{\rho})
      &=-\bar{\lambda}_1\log_2\bar{\lambda}_1-\bar{\lambda}_2\log_2\bar{\lambda}_2\\\nonumber
      &=-\frac{1}{2}[1+\mu(\tau) E(\beta)]\log_2[1+\mu(\tau) E(\beta)]\\
      &-\frac{1}{2}[1-\mu(\tau) E(\beta)]\log_2[1-\mu(\tau) E(\beta)]+1.
\end{align}
Since $\rho_{X_0}(\tau)$ and $\rho_E(0)$ in \eqref{rho_f} are only unitarily related, $\bar{S}(\rho)=S[\rho_E(0)]$. With the eigenvalues of $\rho_E(0)$, $\{[1+E(\beta)]/2,[1-E(\beta)]/2\}$, $\bar{S}$ is time-independent:
\begin{align}\label{S bar}\nonumber
   \bar{S}&=S[\rho_E(0)]\\
  &=-\frac{1}{2}[1+ E(\beta)]\log_2[1+ E(\beta)]\\\nonumber
      &-\frac{1}{2}[1- E(\beta)]\log_2[1-E(\beta)]+1.
\end{align}
The Holevo quantity $\chi(\rho)$ is
\begin{align}\label{Holevo single}\nonumber
    \chi(\rho)&=S(\bar{\rho})-\bar{S}\\\nonumber
      &=-\frac{1}{2}[1+\mu(\tau) E(\beta)]\log_2[1+\mu(\tau) E(\beta)]\\\nonumber
      &-\frac{1}{2}[1-\mu(\tau) E(\beta)]\log_2[1-\mu(\tau) E(\beta)]\\
     &+\frac{1}{2}[1+ E(\beta)]\log_2[1+ E(\beta)]\\
      &+\frac{1}{2}[1- E(\beta)]\log_2[1-E(\beta)].\nonumber
\end{align}
$\chi(\rho)$ in \eqref{Holevo single} can be viewed as a difference between Shannon entropies, $H\{[1+\mu(\tau) E(\beta)]/2\}$ and $H\{[1+E(\beta)]/2\}$ for a qubit. All the relevant parameters are contained in $\mu(\tau)$ and the temperature dependence in $E(\beta)$. The Shannon entropy $H(p)$ for $p\geq1/2$ is a decreasing function of $p$ and its slop is steeper as $p\to1$. As $\mu(\tau)$ gets smaller $\chi(\rho)$ gets bigger. Since $|dH(p)/dp|$ approaches the maximum as $E(\beta)\to1$ ($T\to 0$), with $\mu(\tau)$ being fixed, $\chi(\rho)$ gets bigger as $T\to 0$, i.e. as temperature decreases, information is better transferred. This is intuitively clear and consistent with the results in Refs. \cite{Lee:2024iso, Lee:2023ozm} that the distinguishability decreases as temperature increases. 


\subsection{Short time behavior}
\begin{figure}[t!]
\includegraphics[width=0.9\linewidth, height=4cm]{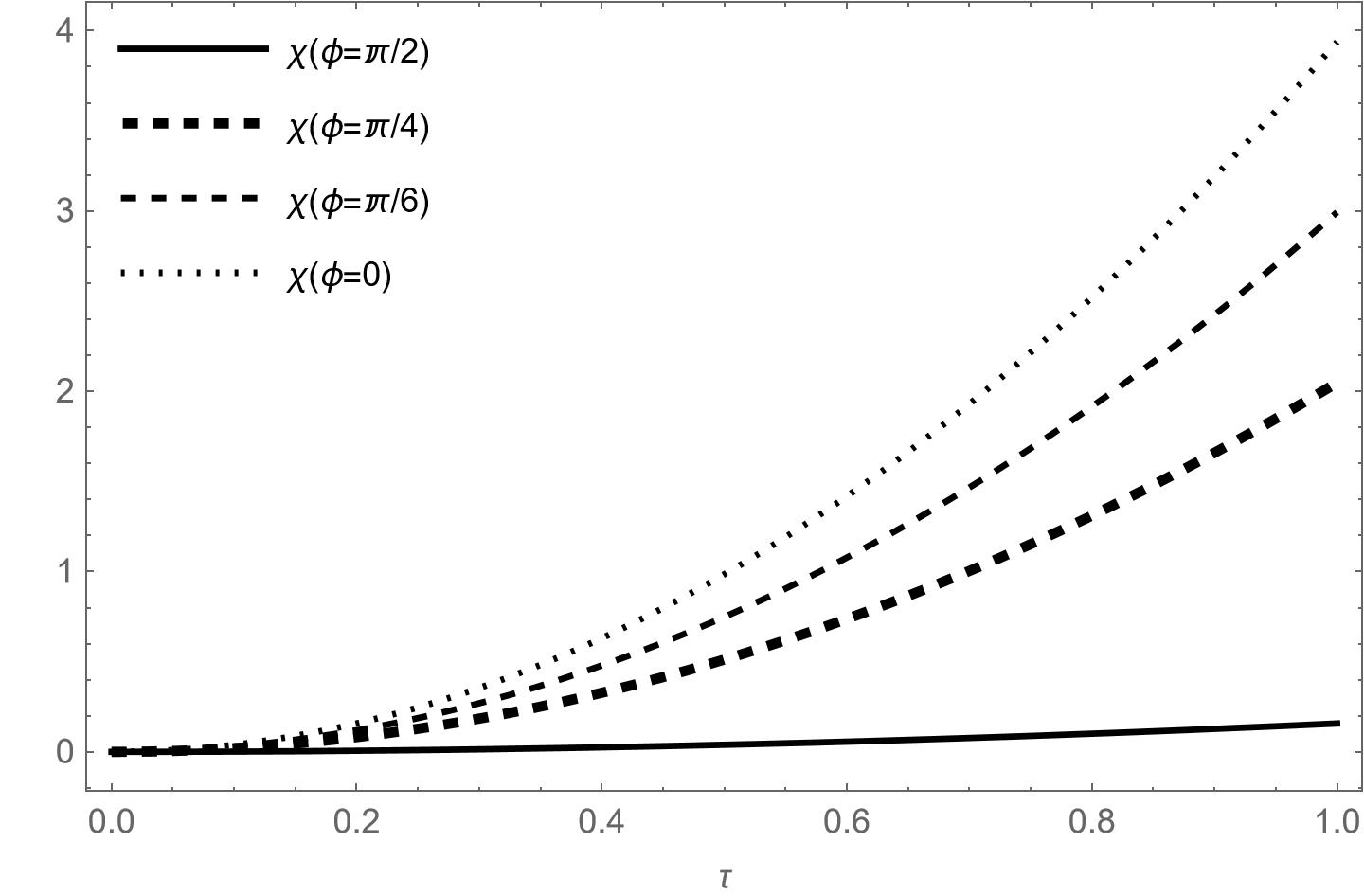} 
\caption{\label{fig:Holevo_shortime_phi} Short-time behaviors of the Holevo quantities at different initial conditions $\phi=(\frac{\pi}{2},\frac{\pi}{4},\frac{\pi}{6},0 )$. ($M=1$. $\Omega=5$, $r=1$, $\theta=0$, $|\alpha|=1$, $\tilde{g}=\frac{1}{2}$, $\Delta=1$, $\beta=10$).}
\end{figure}

As seen in Fig. \ref{fig:Holevo_infinity}, $\chi(\rho)$ quickly grows and gets stabilized from the beginning. To verify this behavior, we investigate a short-time behavior of $\chi(\rho)$. For this purpose, as $\chi(\rho)$ is determined only by $\mu(\tau)$ and $E(\beta)$, we expand $\mu(\tau)$ in $\tau$ and $\tilde{g}$ up to $O(\tau^2)$ and $O(\tilde{g}^2)$ according to the small coupling approximation, $\tilde{g}\ll1$. Following  the details given in Appendix \ref{A:Short time}, $\mu^2(\tau)$ in \eqref{C:mu_expansion} is expressed as 
\begin{align}\label{mu_expansion}\nonumber
  &\mu^2(\tau)\\
   &=1-2\tau^2\eta^2\tilde{g}^2[\cos^2\phi+4\tilde{\Delta}^2\sin^2\phi\cos^2(2q\tilde{g}\sin\phi)]\\\nonumber
   &+O(\tau^3)+O(\tilde{g}^3)
\end{align}
This expression clearly shows that $\mu^2(\tau)$ is a decreasing function in $\tau$ and hence $\chi(\rho)$ is an increasing function in $\tau$, since, as seen in \eqref{Holevo single}, $\chi(\rho)$ is a decreasing function in $\mu(\tau)$. Expanding $\chi(\rho)$ in $\tau$ up to $O(\tau^2)$ and $O(\tilde{g}^2)$ from \eqref{C:chi_O(2)},
\begin{align}\label{chi_O(2)}\nonumber
    \chi(\rho)&=\frac{E^2(\beta)}{4}\{\log_2[1-E(\beta)]-\log_2[1+E(\beta)]\}\mu''(0)\tau^2\\\nonumber
&+O(\tau^3)+O(\tilde{g}^3)\\\nonumber
    &=\frac{E^2(\beta)}{2}\log_2\frac{1+E(\beta)}{1-E(\beta)}\\\nonumber
&\times\tau^2\eta^2\tilde{g}^2[\cos^2\phi+4\tilde{\Delta}^2\sin^2\phi\cos^2(2q\tilde{g}\sin\phi)]\\\nonumber
 &+O(\tau^3)+O(\tilde{g}^3)\\
&=\Lambda\tau^2+O(\tau^3)+O(\tilde{g}^3),
\end{align}
where 
\begin{align}\label{Lambda}
    \Lambda\equiv\frac{E^2(\beta)}{2}\log_2\frac{1+E(\beta)}{1-E(\beta)}[1-(1-4\tilde{\Delta}^2)\sin^2\phi]\eta^2\tilde{g}^2,
\end{align}
and from \eqref{eta(t),q(t)}
\begin{align}\label{eta,q}
&\eta\equiv \eta(0)=\frac{1}{M\Omega}(\cosh{2r}+\cos\theta\sinh{2r}),\\
&q\equiv q(0)=\sqrt{\frac{2}{M\Omega}}|\alpha|\cos\phi.
\end{align}
Equation \eqref{chi_O(2)} is one of our main results. It shows that $\chi(\rho)$ grows quadratically in time for short times. The speed of the growth is given by the factor $\Lambda$ from \eqref{Lambda}. Interestingly, the larger $\phi$ is, the larger the initial growth of  $\chi(\rho)$ is, which is opposite of behavior for $\tau\to\infty$. This is shown in Fig. \ref{fig:Holevo_phi}. Figure \ref{fig:Holevo_shortime_phi} clearly shows those short-time behaviors at different values of $\phi$.


\subsection{Asymptotic Holevo quantity}
Let us introduce the following asymptotic quantity $\chi_{\infty}(\rho)$. Noticing that $\chi(\rho)$ quickly approaches 
$\chi_{\infty}(\rho)$ as it evolves, $\chi_{\infty}(\rho)$ is not only simpler to analyze but also more relevant to the time scale in our focus on classicality. 
\begin{align}\label{def:Holevo single infinite}
    \chi_{\infty}(\rho)&=\lim _{\tau\to\infty}\chi(\rho).
\end{align}
 It turns out that the contribution of $D_{12}(\tau)$ and $D_3(\tau)$ in \eqref{D12,D3} to the Holevo quantity $\chi(\rho)$ in \eqref{Holevo single} with $\mu_i$ in \eqref{mui} becomes negligible for large $\tau$. The difference between the exact expression $\chi(\rho)$ and the asymptotic quantity $\chi_\infty(\rho)$ in \eqref{Holevo single infinite} is negligible after some initial stabilization period, and it is easier to analyze $\chi_\infty(\rho)$ than $\chi(\rho)$.
As $\tau\to \infty$ in \eqref{B:I(dk)},
\begin{align}\label{I_limits}
    \lim_{\tau\to\infty}I[-k,k_-,-k_0]&= \lim_{\tau\to\infty}I[k,k_-,k_0]\\\nonumber
   = \lim_{\tau\to\infty}I[-k,k_+,-k_0]&=\lim_{\tau\to\infty}I[k,k_+,k_0]=0.
\end{align}
Especially, it is easy to notice in \eqref{D12,D3} that $D_{12}(\tau)$ and $\text{Re}[D_3(\tau)]$ at $\phi=0$ identically vanish, without making any contribution to $\mu_i(\tau)$ in \eqref{mui}, i.e., there is no difference between $\chi(\rho)$ and $\chi_\infty(\rho)$ at $\phi=0$,
\begin{align}\label{phi0_chi=chi_inf}
\chi(\rho)=\chi_\infty(\rho)~\text{at}~\phi=0.
\end{align}
This implies that at $\phi=0$, $\chi(\rho)$ has only an oscillatory pattern without the asymptotic behavior, shown in Fig. \ref{fig:Holevo_phi}.
\begin{figure}[t!]
\includegraphics[width=0.9\linewidth, height=4cm]{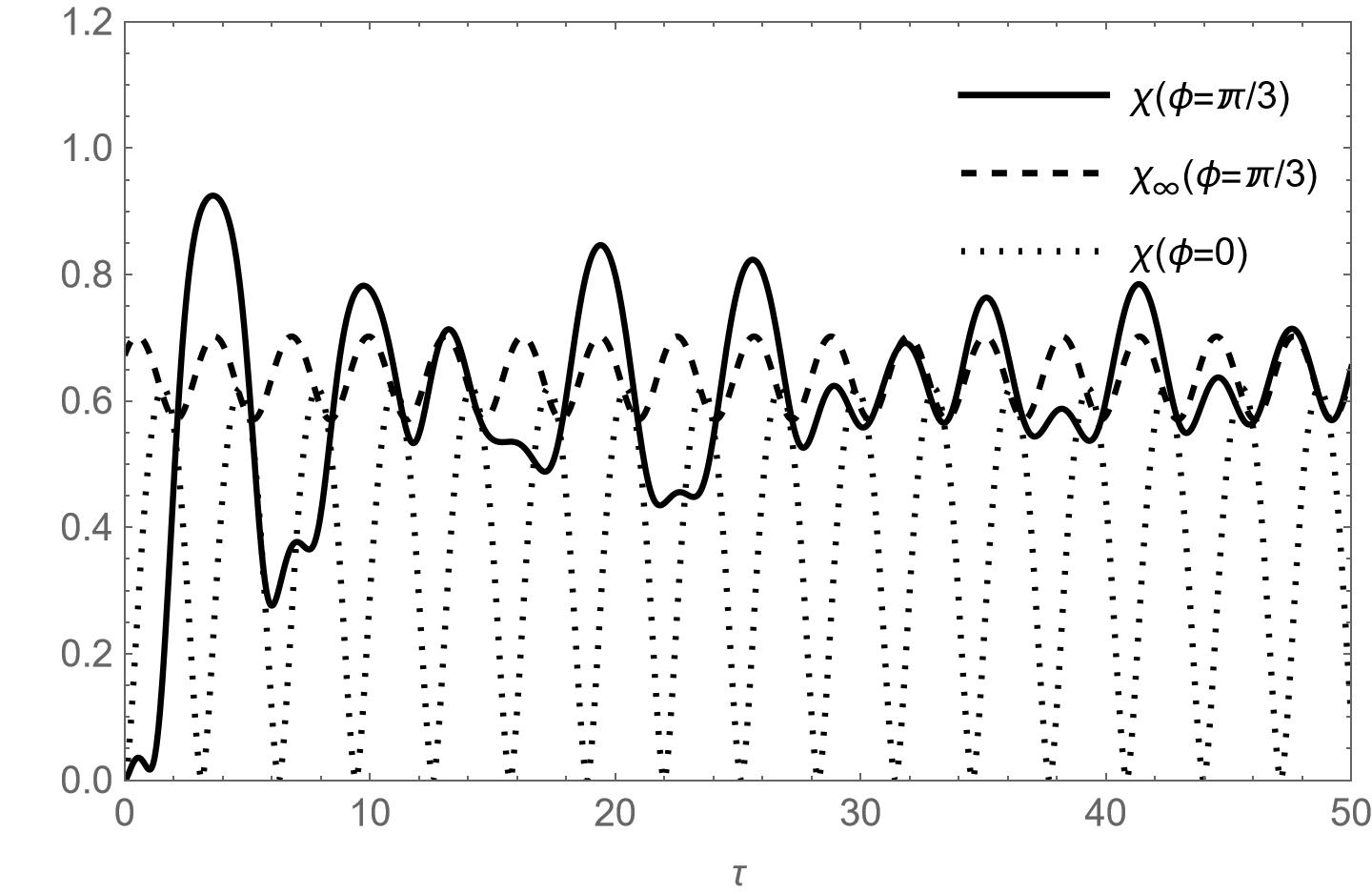} 
\caption{\label{fig:Holevo_infinity} Comparison of the Holevo quantity $\chi(\rho)$ with the asymptotic form $\chi_\infty(\rho)$ ($M=1$. $\Omega=5$, $r=1$, $\theta=0$, $|\alpha|=1$, $\phi=\frac{\pi}{3}$, $\tilde{g}=\frac{1}{2}$, $\Delta=1$, $\beta=10$).}
\end{figure}
In \eqref{D12,D3} $D_{12}(\tau)$ and $D_3(\tau)$ asymptotically vanish as $\tau\to\infty$,
\begin{align}\label{D_limits}
\lim_{\tau\to\infty}D_{12}(\tau)=\lim_{\tau\to\infty}D_3(\tau)=0.
\end{align}
The reason that $D_{12}(\tau)$ and $D_3(\tau)$ vanish for large-time scale is due to the self-Hamiltonian $-\Delta \sigma_x/2$ in \eqref{H_S,H_E,H_int}, i.e., nonzero $k=-2\tilde{g}^2\tau\tilde{\Delta}$, which is proportional to $\tau$. For $\phi=0$ there is no $\tilde{\Delta}$ dependence for $\chi(\rho)$, so $\chi(\rho)$ shows a monotonous periodic pattern without information growth. This verifies that the $\phi=0$ case does not show the distinguishability in objectivity shown in Ref. \cite{Lee:2024iso}.
Due to \eqref{D_limits} $\mu_i(\tau)$ tends to approach as
\begin{align}\label{mu_limits}\nonumber
\lim_{\tau\to\infty} \mu_1(\tau)&= \frac{1}{2}\text{Re}(I[0,k_-,0]+I[0,k_+,0])\\
\lim_{\tau\to\infty} \mu_2(\tau)&= \frac{1}{2}\text{Im}(I[0,k_-,0]+I[0,k_+,0])\\
\lim_{\tau\to\infty} \mu_3(\tau)&=0.\nonumber
\end{align}
Thus, the asymptotic quantity $\mu_\infty$ is expressed as
\begin{align}\label{mu_infinity}\nonumber
\mu_\infty&\equiv\lim_{\tau\to\infty}\sqrt{\mu^2_1(\tau)+\mu^2_2(\tau)+\mu^2_3(\tau)}\\\nonumber
&=\frac{1}{2}\left|\mathit{e}^{-\frac{\eta k^2_-}{4}}+\mathit{e}^{-\frac{\eta k^2_+}{4}}e^{iq\delta k_\pm}\right|\\
&=\frac{1}{2}\left\{\mathit{e}^{-\frac{\eta k^2_-}{2}}+\mathit{e}^{-\frac{\eta k^2_+}{2}}\right.\\\nonumber
    &\left.+2\mathit{e}^{-\frac{\eta (k^2_-+k^2_+)}{4}}\cos[2\tilde{g}q_0\sin2\phi]\right\}^{1/2}\\\nonumber
    &\leq \cos[\tilde{g}q_0\sin2\phi]\nonumber,
\end{align}
where $q_0\equiv|\alpha|\sqrt{\frac{2}{M\Omega}}$; $\eta$ and $q$ were defined in \eqref{eta,q}. Figure \ref{fig:Holevo_infinity} shows how quickly $\chi(\rho)$ approaches the corresponding asymptotic limit $\chi_\infty(\rho)$. 

For large $\tau$, the eigenvalues of $\bar{\rho}$, $\bar{\lambda}_{1,2}$, approach
\begin{align}\label{lambda bar}\nonumber
    \bar{\lambda}_1&\to\frac{1}{2}[1+\mu_{\infty} E(\beta)],\\
    \bar{\lambda}_2&\to\frac{1}{2}[1-\mu_{\infty} E(\beta)]
\end{align}
and $S(\bar{\rho})$ approaches $S_{\infty}(\bar{\rho})$,
\begin{align}\label{S rho bar}\nonumber
    S_{\infty}(\bar{\rho})
      &=-\frac{1}{2}[1+\mu_\infty E(\beta)]\log_2[1+\mu_\infty E(\beta)]\\
      &-\frac{1}{2}[1-\mu_\infty E(\beta)]\log_2[1-\mu_\infty E(\beta)]+1.
\end{align}
The asymptotic Holevo quantity $\chi_{\infty}(\rho)$ is
\begin{align}\label{Holevo single infinite}\nonumber
    \chi_{\infty}(\rho)&=S_{\infty}(\bar{\rho})-\bar{S}\\\nonumber
      &=-\frac{1}{2}(1+\mu_{\infty})\log_2[1+\mu_{\infty} E(\beta)]\\\nonumber
      &-\frac{1}{2}[1-\mu_{\infty} E(\beta)]\log_2[1-\mu_{\infty} E(\beta)]\\
     &+\frac{1}{2}[1+ E(\beta)]\log_2[1+ E(\beta)]\\\nonumber
      &+\frac{1}{2}[1- E(\beta)]\log_2[1-E(\beta)].\nonumber
\end{align}
\subsection{Parameter dependence}
\begin{figure}[t!]
\includegraphics[width=0.9\linewidth, height=4cm]{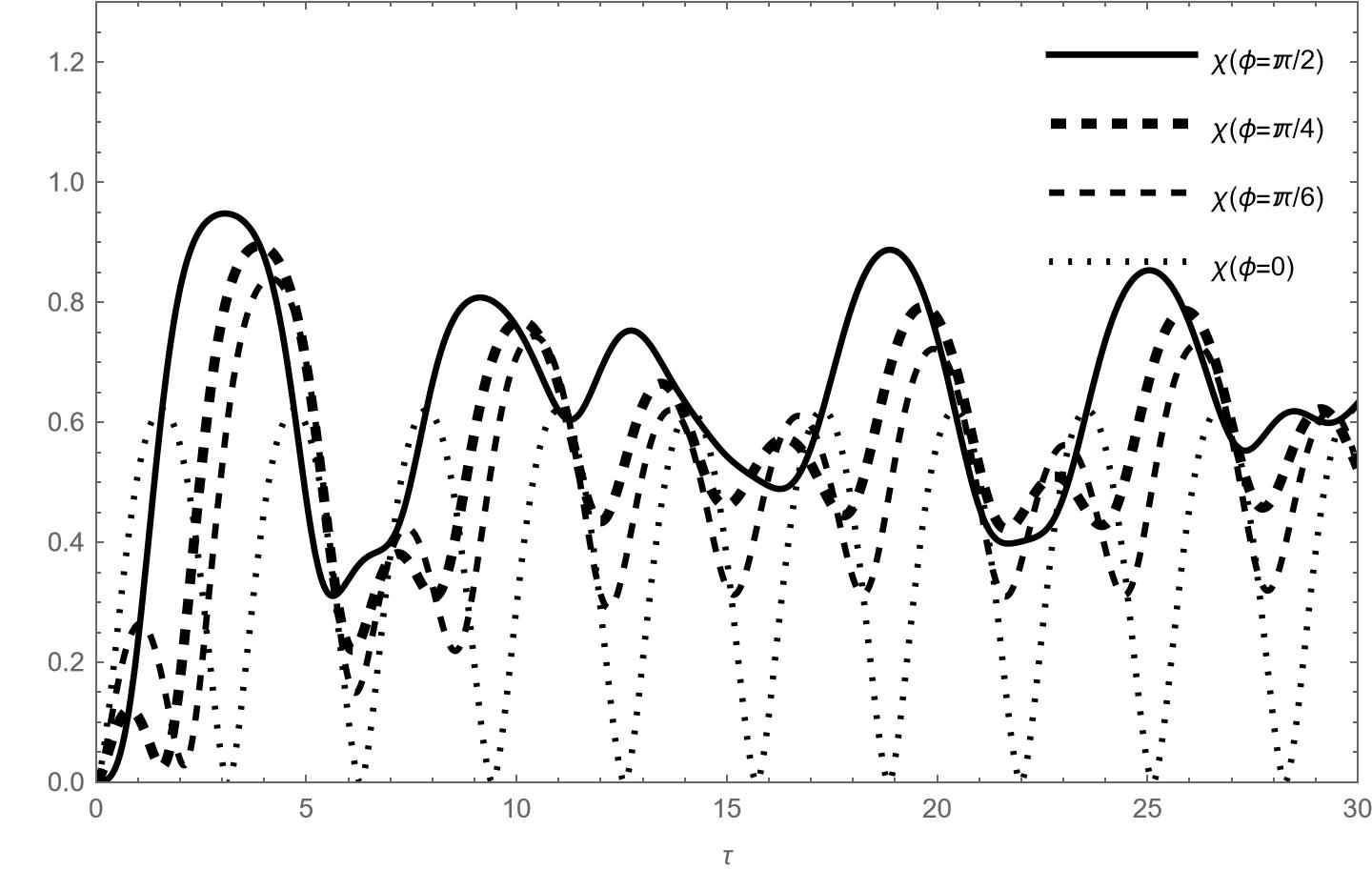} 
\caption{\label{fig:Holevo_phi} The Holevo quantities at different initial conditions, $\phi=(\frac{\pi}{2},\frac{\pi}{4},\frac{\pi}{6},0 )$. ($M=1$. $\Omega=5$, $r=1$, $\theta=0$, $|\alpha|=1$, $\tilde{g}=\frac{1}{2}$, $\Delta=1$, $\beta=10$).}
\end{figure}

\begin{figure}[t!]
\includegraphics[width=0.9\linewidth, height=4cm]{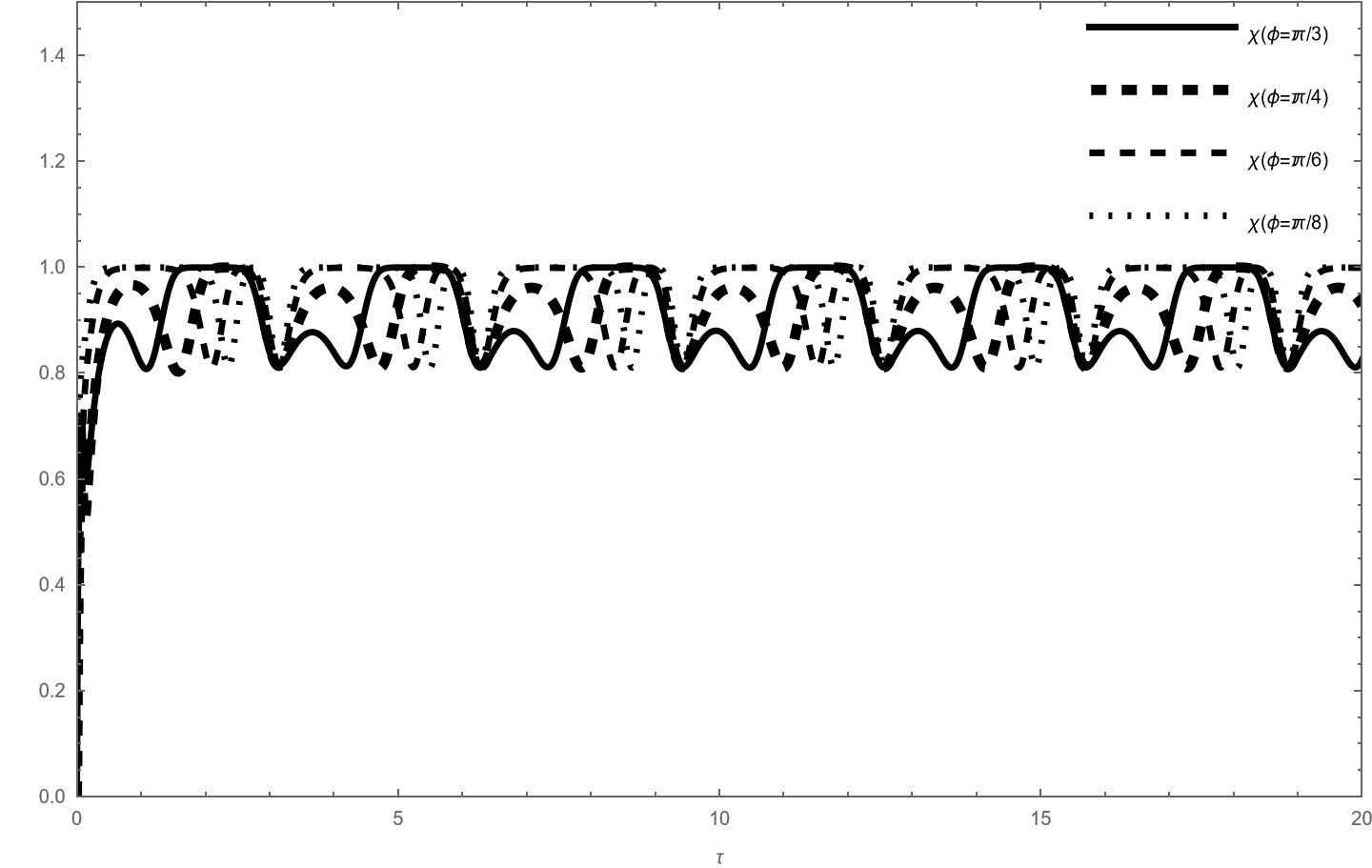} 
\caption{\label{fig:Holevo_phi_max} The Holevo quantities reaching the maxima with the squeezing parameter $\eta$ fixed and $q$ varied at $\phi=(\frac{\pi}{2},\frac{\pi}{4},\frac{\pi}{6},0)$ . ($M=1$, $r=1$, $\theta=0$, $|\alpha|=1$, $\Omega$ under the condition $\psi=2\tilde{g}q_0\sin2\phi=\pi$, $\tilde{g}=\frac{1}{2}$, $\Delta=1$, $\beta=10$).}
\end{figure}

\begin{figure}[t!]
\includegraphics[width=0.9\linewidth, height=4cm]{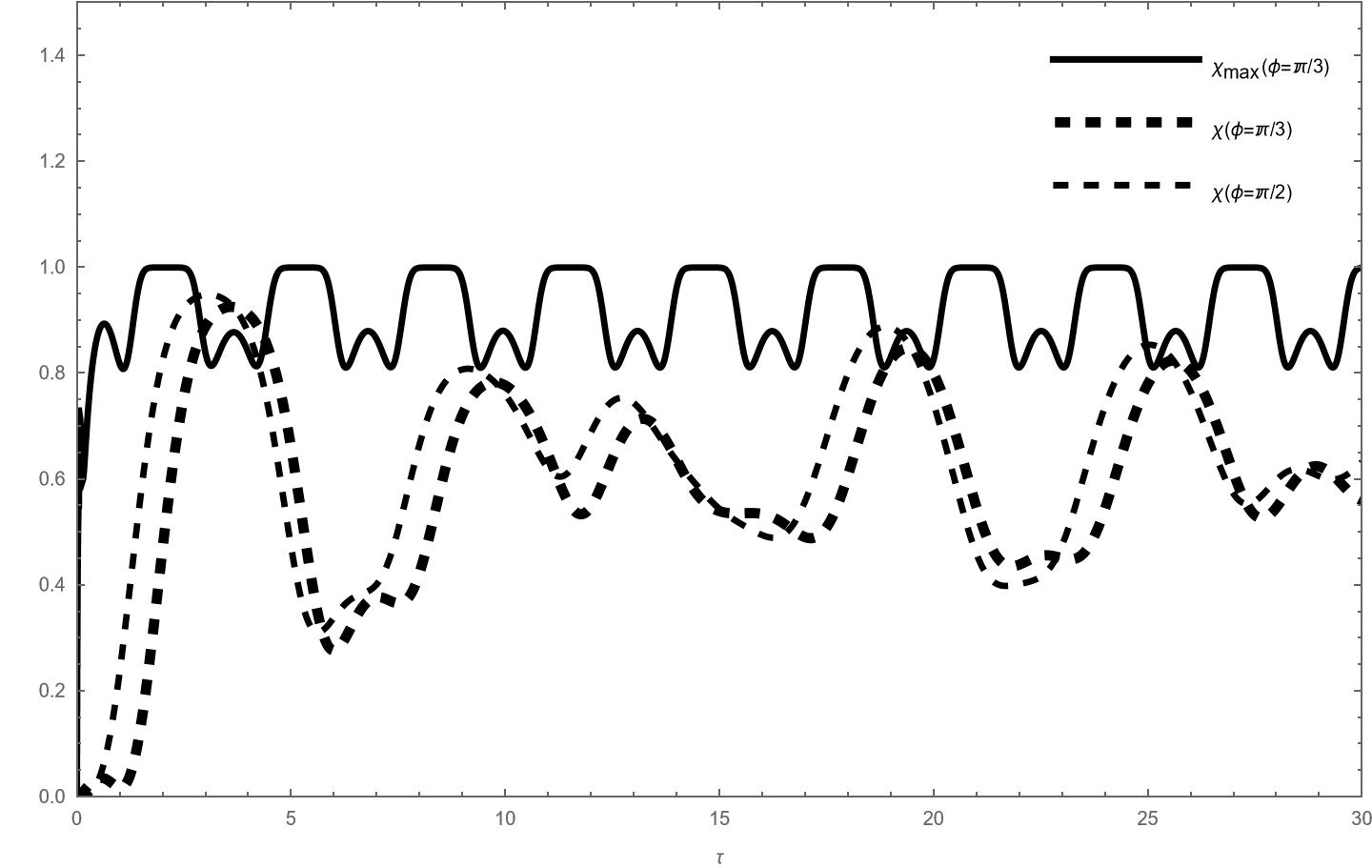} 
\caption{\label{fig:Holevo_phi_max_comp} Comparison among the Holevo quantities at $\psi=2\tilde{g}q_0\sin2\pi/3=\pi$, $\frac{\pi}{3}$ and $\frac{\pi}{2}$ with $\Omega=5$ . ($M=1$, $r=1$, $\theta=0$, $|\alpha|=1$, $\tilde{g}=\frac{1}{2}$, $\Delta=1$, $\beta=10$).}
\end{figure}

\begin{figure}[t!]
\includegraphics[width=0.9\linewidth, height=4cm]{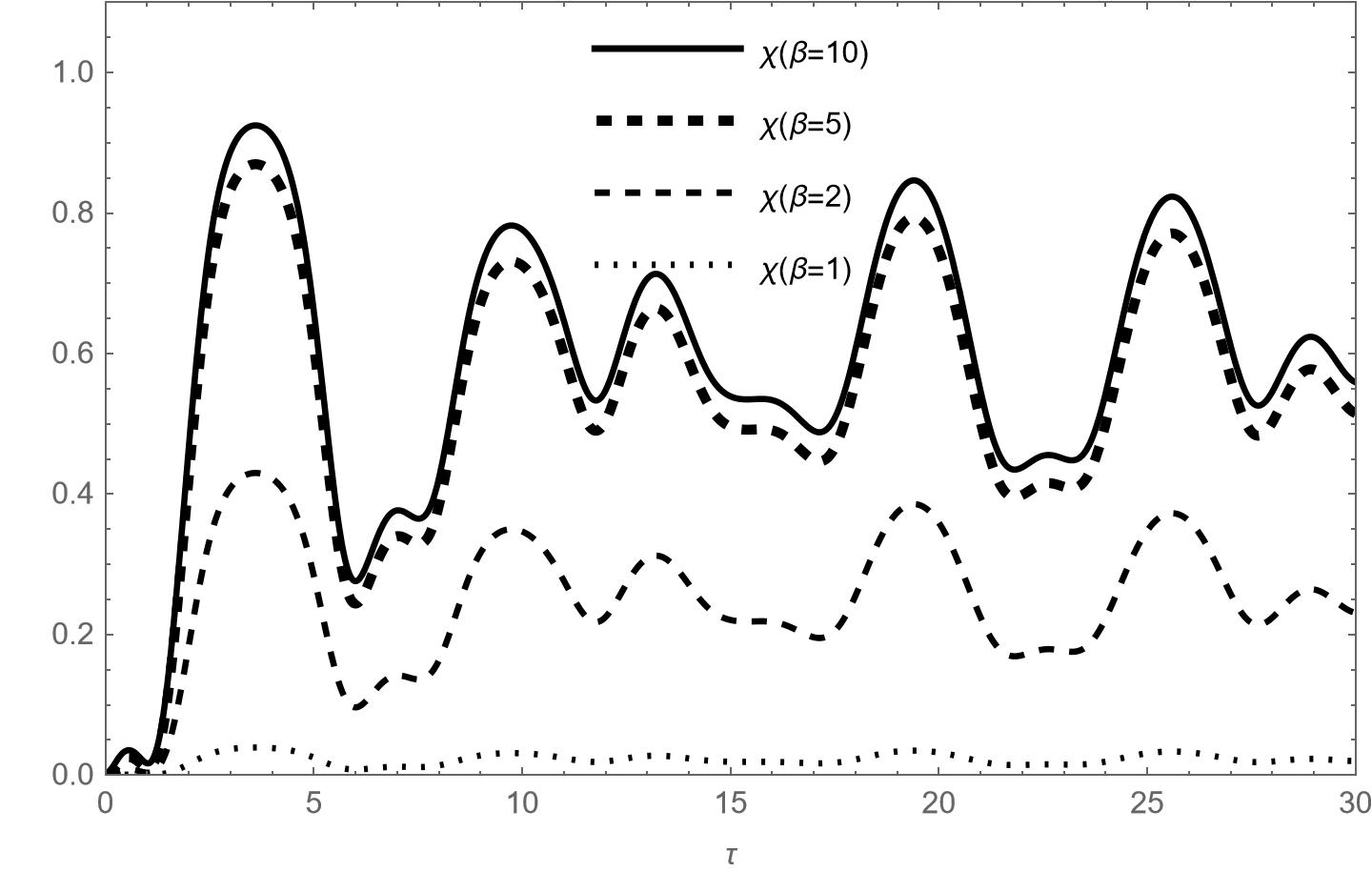} 
\caption{\label{fig:Holevo_beta} Temperature dependence of the Holevo quantities at $\beta=(10,5,2,1)$ . ($M=1$. $\Omega=5$, $r=1$, $\theta=0$, $|\alpha|=1$, $\phi=\frac{\pi}{3}$, $\tilde{g}=\frac{1}{2}$, $\Delta=1$).}
\end{figure}

$\chi(\rho)$ depends on parameters from the Hamiltonian, $(\Delta,g,\Omega,M)$ and the initial conditions, $(\beta,\phi,\eta,q_0)$. For our interest in large-time behaviors, we analyze $\chi_\infty(\rho)$ instead of $\chi(\rho)$. $\chi_\infty(\rho)$ is parametrized by two parameters, $\mu_\infty$ and $E(\beta)$. It increases as $\mu_\infty$ as well as $E(\beta)$ decreases. $\mu_\infty$ in \eqref{mu_infinity} is a function of $|\vec{\gamma}|$, where
$\vec{\gamma}\equiv \vec{\alpha}+\vec{\beta}$, 
whose magnitudes and the relative angle $\psi$ are given by 
\begin{align}\label{alpha,beta,psi}
    |\vec{\alpha}|(\tilde{g},\eta,\phi,\tau)&\equiv\mathit{e}^{-\frac{\eta k^2_-}{4}}=\mathit{e}^{-\eta\tilde{g}^2\cos^2(\tau/2+\phi)\sin^2(\tau/2)},\\
    |\vec{\beta}|(\tilde{g},\eta,\phi,\tau)&\equiv\mathit{e}^{-\frac{\eta k^2_+}{4}}=\mathit{e}^{-\eta\tilde{g}^2\sin^2(\tau/2+\phi)\cos^2(\tau/2)},\\
    \psi(\tilde{g},q_0,\phi)&\equiv 2\tilde{g}q_0\sin2\phi.
\end{align}
$|\vec{\alpha}|$ and $|\vec{\beta}|$ depend on a squeezing parameter $\eta$, $\phi$ and $\tau$ contained in $k_\pm$ in \eqref{k}.
This geometrical description is useful to analyze $\mu_\infty(|\vec{\gamma}|)$ and hence $\chi_\infty(\rho)$. It is immediately noticeable that  for large $\eta$,  $\chi_\infty(\rho)$ is high.
We list possible cases when  $\chi_\infty(\rho)$ is minimized and maximized with the configuration of ($|\vec{\alpha}|, |\vec{\beta}|,\psi)$ in \eqref{alpha,beta,psi} for fixed temperature first.\\
Maximizing $\chi(\rho)$ ($\mu_\infty\to0$):
\begin{enumerate}
    \item  $\eta\to\infty$ ($\vec{\alpha}=\vec{\beta}=0$): This corresponds to the uniform distribution of $p(X_0)$. It leads to $S(\bar{\rho})\to 1$ and $\chi(\rho)=\chi_M(\beta)$.
    \item $|\vec{\alpha}|=|\vec{\beta}|\to$ $\tau+\phi=(2n+1)\pi$ $(n\in\mathbb{Z})$ and $\psi=(2n+1)\pi$: As long as $\psi=(2n+1)\pi$ is satisfied, $\chi_\infty(\rho)$ arrives at the local maxima at $\tau+\phi=(2n+1)\pi$.
    \item $\tilde{g}\uparrow$ $\rightarrow$ $ |\vec{\alpha}|\downarrow, |\vec{\beta}|\downarrow$ and $\psi\uparrow$: Increasing $\tilde{g}$ increases $\chi_{\infty}(\rho)$
    \item $\phi\uparrow$: Apart from oscillatory parts, as $\phi$ increases, $\chi_\infty(\rho)$ increases.
\end{enumerate}
Figure \ref{fig:Holevo_phi} shows that the larger $\phi$ is, the more information is encoded in the spins. On the other hand, Fig. \ref{fig:Holevo_phi_max} shows that when $\vec{\alpha}$ and $\vec{\beta}$ are antiparallel, $\chi_\infty(\rho)$ is maximized, i.e., for the parameters $\tilde{g}$, $q$, and $\phi$ satisfying 
\begin{align}\label{maximization}
    2\tilde{g}|\alpha|\sqrt{\frac{2}{M\Omega}}\sin2\phi=\pi,
\end{align}
all $\chi(\rho)$ quickly get maximized regardless of values of $\phi$. $\chi(\rho)$ with this condition is the maximum rather than at $\phi=\pi/2$. Figure \ref{fig:Holevo_phi_max_comp} shows that $\chi(\rho)$ under the maximization condition \eqref{maximization} asymptotically has a higher value larger than those without it. Also, it shows that nonzero $\phi$ is crucial to have an asymptotic behavior stabilizing the maximum.\\
Minimizing $\chi(\rho)$ ($\mu_\infty\to1$):
\begin{enumerate}
    \item $\eta\to0$ ($\vec{\alpha}=\vec{\beta}\to1$) and $\psi=0$: This corresponds to $p(X_0)\to\delta(X_0-q)$, i.e. the initial density matrix is a localized pure state, which leads to $\chi_\infty(\rho)=0$. There are two cases for $\psi=0$.
    \item
    $\sin2\phi=0$: As 
 seen in \eqref{alpha,beta,psi}, at $\phi=0$, i.e., $|\vec{\alpha}|=|\vec{\beta}|$, $\chi(\rho)$ vanishes, while at $\phi=\pi/2$, it does not since $|\vec{\alpha}|\neq |\vec{\beta}|$. Thus, the $\phi=0$ case is consistent with the result that $\sin\phi=0$ does not lead to a vanishing generalized overlap (distinguishability) \cite{Lee:2024iso}.
\end{enumerate}
Now we confirm our intuition that reducing temperature increases distinguishability. As mentioned below \eqref {Holevo single}, $\chi_\infty(\rho)$ is a difference between the Shannon entropies $H\{[1+\mu_\infty E(\beta)]/2\}$ and $H\{[1+E(\beta)]/2\}$. This difference gets larger as $E(\beta)$ is closer to 1. Recognizing that for $p>1/2$, $dH(p)/dp<0$, and $|dH(p)/dp|$ increases as $p$ increase. For $\beta'>\beta$ $(T>T')$, 
\begin{align}\label{Holevo_beta_diff}\nonumber
    &\chi_\infty(\beta')-\chi_\infty(\beta)\\\nonumber
    &=H\{[1+\mu_\infty E(\beta')]/2\}-H\{[1+\mu_\infty E(\beta)]/2\}\\\nonumber
    &-(H\{[1+ E(\beta')]/2\}-H\{[1+ E(\beta)]/2\})\\
&\approx\Delta E\left[\mu_\infty\frac{\partial H(p')}{\partial p'}-\frac{\partial H(p)}{\partial p}\right]>0,
\end{align}
where $p'=[1+\mu_\infty E(\beta)]/2$ and  $p=[1+ E(\beta)]/2$.
This means that as a temperature gets lower $\chi_\infty(\rho)$ gets larger. This is consistent with the fact that lowering temperature enhances the distinguishability \cite{Lee:2024iso}. 
 All theses observations are consistent with effects on the objectivity in Ref. \cite{Lee:2024iso}. The temperature dependence of the Holevo quantity is shown in Fig. \ref{fig:Holevo_beta}.

 Finally, we wish to mention the maximum $\chi(\rho)$. Define
\begin{align}\label{g_inequality}\nonumber
    &g[\mu(\tau) E(\beta)]\\
    &\equiv-\frac{1}{2}[1+\mu(\tau) E(\beta)]\log_2[1+\mu(\tau) E(\beta)]\\
      &-\frac{1}{2}[1-\mu(\tau) E(\beta)]\log_2[1-\mu(\tau) E(\beta)]\leq 0.\nonumber
\end{align}
Since $dg(\mu)/d\mu \leq 0$, at $\mu=0$, $ g[\mu (\tau) E(\beta)]=0$, which is the maximum. As $\chi(\rho)=g[\mu(\tau) E(\beta)]-g[E(\beta)]$, $\chi(\rho)$ is the maximum $\chi_M(\beta)$ at $\mu=0$ with $\beta$ fixed.
\begin{align}\label{Holevo_max}\nonumber
    \chi_M(\beta)&\equiv\frac{1}{2}[1+ E(\beta)]\log_2[1+ E(\beta)]\\
   &+\frac{1}{2}[1- E(\beta)]\log_2[1-E(\beta)]\\
   &=-g[E(\beta)]\geq0.\nonumber
\end{align}
Since $-dg[E(\beta)]/dE(\beta) \geq 0$, $\chi_M(\beta)$ is 1, the largest, at $E(\beta)=1$, i.e. $\beta\to \infty$ $(T=0)$.
\begin{align}\label{Holevo_range}
 0\leq\chi(\rho)\leq \chi_M(\beta)\leq 1.
\end{align}
\section{Conclusion}
This article examined how the information of continuous degrees of freedom is encoded into a system of finite degrees of freedom with a simple system, a boson-spin model. On the dualism of the quantum Darwinism between objectification and information transfer, we wanted to verify the objectivity that was investigated with the decoherence factor for decoherence and the generalized overlap for distinguishability in the last work \cite{Lee:2024iso} for the boson-spin system, especially distinguishability, with a point of view of information transfer by calculating the mutual information bound, the Holevo quantity. 

We considered the simplest situation, in which a central harmonic oscillator and a single spin system remain after unobserved spins traced out. In order to apply for the Holevo theorem, we assumed that the system is in a perfect decoherence in a position basis up to a local unitary transformation.    
An initial probability distribution is chosen from a displacement and squeezed state and an initial state of a spin is chosen to be a thermal state. In this setting we investigated how the Holevo quantity for our model is shown to be consistent with distinguishability measured by the generalized overlap.
We confirmed that as expected, the high temperature is against the precision of encoding, which is consistent with the result \cite{Lee:2024iso, Lee:2023ozm}. 

We established the role of a self-Hamiltonian of a spin environment, that it is necessary to make sure asymptotic stabilization in the Holevo quantity.
We found the relation between the Holevo quantity and the parameters in the system. We confirmed the intuitive relations of the Holevo quantity with a squeezing parameter, temperature and a coupling constant, as increasing a squeezing parameter and a coupling constant and decreasing temperature increase the Holevo quantity. Increasing displacement parameter $q$ also increases the Holevo quantity, which can be understood as $q$ is the size of position space in statistical distribution. Especially, we showed why a vanishing initial phase $\phi$ in the oscillator trajectory does not provide the distinguishability. This can be an alternative explanation why $\phi=0$ does not provide the objectivity on a position basis in the previous work \cite{Lee:2024iso}.   

The next logical step would be to consider bigger groups of environmental spins and calculate the Holevo quantity for the corresponding one-to-many broadcasting. This presents some technical challenges when calculating the entropy of the average state but we believe those problems can be overcome. 
\begin{acknowledgments}
T.-H.L. and J.K.K. acknowledge the support
from the Polish National Science Center (NCN), Grant No.  2019/35/B/ST2/01896.
\end{acknowledgments}

\appendix
\label{appendix}
\section{Bloch Representation}
Before integrating to get $\mu_i(\tau)$ in \eqref{mui}, we first need to express $\rho_{X_0}(\tau)$ defined in \eqref{rho_f} in the Bloch representation. Given the expression for $U(X_0,t)$ in \eqref{U_Bloch} and an initial thermal state $\rho_E(0)=[\mathbb{I}+E(\beta)\sigma_x]/2$ in \eqref{thermal state}, $\rho_{X_0}(\tau)$ is written in the Bloch form,
\begin{align}\label{A:rho_f}\nonumber
\rho_{X_0}(\tau)&=U(X_0,t)\rho_E(0) U^\dagger(X_0,t)\\
    &=\frac{1}{2}[\mathbb{I}+\vec{a}(X_0,\tau)\cdot\vec{\sigma}],
\end{align}
where
\begin{align}\label{A:ai}\nonumber
     a_1(X_0,\tau)&=E(\beta)(2U_0^2+2U_1^2-1),\\
      a_2(X_0,\tau)
      &=2E(\beta)(U_1U_2-U_0U_3),\\
       a_3(X_0,\tau)
      &=2E(\beta)(U_0U_2+U_1U_3)\nonumber
\end{align}
and from \eqref{U_i}
\begin{align}\label{A:U_i}\nonumber
 U_0&=\cos[\xi(\sin(\tau+\phi)-\sin\phi)]\cos[\tilde{\Delta}(1-\xi^2)\tau]\\\nonumber
U_1&=\cos[\xi(\sin(\tau+\phi)+\sin\phi)]\sin[\tilde{\Delta}(1-\xi^2)\tau]\\
U_2&=\sin[\xi(\sin(\tau+\phi)+\sin\phi)]\sin[\tilde{\Delta}(1-\xi^2)\tau]\\
 U_3&=-\sin[\xi(\sin(\tau+\phi)-\sin\phi)]\cos[\tilde{\Delta}(1-\xi^2)\tau].\nonumber
\end{align}
It would be more conveniently to express $a_i(X_0,\tau)$ in \eqref{A:ai} as an exponential series for the Gaussian integration, 
\begin{align}\label{A:ai_exp}\nonumber
      a_1(X_0,\tau)&=E(\beta)\text{Re}[c_{12}(X_0,\tau)],\\
    a_2(X_0,\tau)&=E(\beta)\text{Im}[c_{12}(X_0,\tau)],\\
    a_3(X_0,\tau)&=E(\beta)\text{Re}[c_{3}(X_0,\tau)],\nonumber
\end{align}
where
\begin{align}\label{A:c_functions}\nonumber
c_{12}(X_0,\tau)&\equiv\frac{1}{4}[2\mathit{e}^{ik_-X_0}+2\mathit{e}^{ik_+X_0}\\\nonumber
    &+\mathit{e}^{i(k_-X_0-k X^2_0-k_0)}+\mathit{e}^{i(k_-X_0+k X^2_0+k_0)}\\\nonumber
    &-\mathit{e}^{i(k_+X_0-k X^2_0-k_0)}-\mathit{e}^{i(k_+X_0+k X^2_0+k_0)}],\\
    c_{3}(X_0,\tau)&\equiv\frac{1}{2}[\mathit{e}^{i[-(k_+-k_-)X_0/2+k X^2_0+k_0]}\\
    &-\mathit{e}^{i[(k_+-k_-)X_0/2+k X^2_0+k_0]}],\nonumber
\end{align}
with $\tilde{g}\equiv g/\Omega$ and
\begin{align}\label{A:k}\nonumber
    &k_-=2\tilde{g}[\sin(\tau+\phi)-\sin\phi],\\\nonumber
     &k_+=2\tilde{g}[\sin(\tau+\phi)+\sin\phi],\\
     &k=-2\tilde{g}^2\tau\tilde{\Delta},\\
     &k_0=2\tau\tilde{\Delta}.\nonumber
\end{align}

\section{Gaussian integrals}\label{A:gaussian}
The Holevo quantity $\chi(\rho)$ consists of two parts $S(\bar{\rho})$ and $\bar{S}$ in \eqref{Holevo}. Computing $S(\bar{\rho})$ requires the Gaussian integral with the initial probability distribution in \eqref{average rho}. The Gaussian distribution $p(X_0)$ is given in \eqref{p(x0)},
\begin{align}\label{B:p(x0)}
    p(X_0)=\left(\frac{1}{\pi\eta}\right)^{1/2} \exp\left[-\frac{(X_0-q)^2}{\eta}\right],
\end{align}
where $(\eta, q)$ were defined in \eqref{eta,q}.
Since $p(X_0)$ in \eqref{p(x0)} and $(c_{12},c_3)$ are  Gaussian in complex space, from \eqref{A:ai_exp} and \eqref{A:c_functions} the integrals $\mu_i(\tau)$ in \eqref{mui} can be written as Gaussian integrals 
\begin{align}\label{B:mu_i}\nonumber
    \mu_1(\tau)&=\text{Re}\int dX_0p(X_0)c_{12}(X_0,\tau),\\
    \mu_2(\tau)&=\text{Im}\int dX_0p(X_0)c_{12}(X_0,\tau),\\
    \mu_3(\tau)&=\text{Re}\int dX_0p(X_0)c_{3}(X_0,\tau).\nonumber
\end{align}
Using the following integral formula
\begin{align}\label{B:integral formula}
    I[a,b,c]&=\sqrt{\frac{1}{\pi\eta}}\int^\infty_{-\infty}dy \mathit{e}^{-(y-y_0)^2/\eta+iay^2+iby+ic}\\
    &=\sqrt{\frac{1}{1-ia\eta}}\exp\left[\frac{ia}{1-ia\eta}\left(y^2_0+\frac{b}{a}y_0+i\frac{b^2\eta}{4a}\right)+ic\right],\nonumber
\end{align}
$\mu_i(\tau)$ in \eqref{B:mu_i} are obtained as
\begin{align}\label{B:integral_c12}\nonumber
    &\int dX_0 p(X_0)c_{12}(X_0,\tau)\\\nonumber
    &=\frac{1}{4}[2I[0,k_-,0]+2I[0,k_+,0]\\
    &+I[-k,k_-,-k_0]+I[k,k_-,k_0]\\
    &-I[-k,k_+,-k_0]-I[k,k_+,k_0],\nonumber
\end{align}
and
\begin{align}\label{B:integral_c3}\nonumber
     &\int dX_0 p(X_0,\tau)c_{3}(X_0,\tau)\\
     &=\frac{1}{2}(I[k,-\delta k/2,k_0]-I[k,\delta k/2,k_0]),
\end{align}
where
\begin{align}\label{B:I(k)}\nonumber
   &I[0,k_-,0]=\mathit{e}^{-\frac{\eta k^2_-}{4}}\mathit{e}^{ik_-q},\\
&I[0,k_+,0]=\mathit{e}^{-\frac{\eta k^2_+}{4}}\mathit{e}^{ik_+q},\\\nonumber
     &I[-k,k_-,-k_0]=\sqrt{\frac{1}{1+ik\eta}}\mathit{e}^{\frac{-ik}{1+ik\eta}\left(q^2-\frac{k_-}{k}q-i\frac{k^2_-\eta}{4k}\right)-ik_0},\\ \nonumber 
     &I[k,k_-,k_0]=\sqrt{\frac{1}{1-ik\eta}}\mathit{e}^{\frac{ik}{1-ik\eta}\left(q^2+\frac{k_-}{k}q+i\frac{k^2_-\eta}{4k}\right)+ik_0},\\\nonumber
       &I[-k,k_+,-k_0]=\sqrt{\frac{1}{1+ik\eta}}\mathit{e}^{\frac{-ik}{1+ik\eta}\left(q^2-\frac{k_+}{k}q-i\frac{k^2_+\eta}{4k}\right)-ik_0},\\\nonumber
&I[k,k_+,k_0]=\sqrt{\frac{1}{1-ik\eta}}\mathit{e}^{\frac{ik}{1-ik\eta}\left(q^2+\frac{k_+}{k}q+i\frac{k^2_+\eta}{4k}\right)+ik_0},
\end{align}
and with $\delta k\equiv k_+-k_-$
\begin{align}\label{B:I(dk)}\nonumber
   &I[-k,\delta k/2,-k_0]=\sqrt{\frac{1}{1+ik\eta}}\mathit{e}^{\frac{-ik}{1+ik\eta}\left(q^2-\frac{\delta k_\pm}{2k}q-i\frac{\delta k_\pm^2\eta}{16k}\right)-ik_0},\\
    &I[k,\delta k/2,k_0]=\sqrt{\frac{1}{1-ik\eta}}\mathit{e}^{\frac{ik}{1-ik\eta}\left(q^2+\frac{\delta k}{2k}q+i\frac{\delta k^2\eta}{16k}\right)+ik_0}.
\end{align}
The decaying quantities $D_{12}(\tau)$ and $D_3(\tau)$ as $t\to\infty$ are defined as
\begin{align}\label{B:D12,D3}\nonumber
D_{12}(\tau)&\equiv \frac{1}{2}(I[-k,k_-,-k_0]+I[k,k_-,k_0]\\
&-I[-k,k_+,-k_0]-I[k,k_+,k_0]),\\
D_3(\tau)&\equiv \frac{1}{2}(I[-k,\delta k/2,-k_0]-I[k,\delta k/2,k_0]).\nonumber
\end{align}
\section{Short time expansion}\label{A:Short time}
We expand $I[\cdot,\cdot,\cdot]$ defined in \eqref{B:I(k)} and \eqref{B:D12,D3} up to $O(\tau^2)$ and $O(\tilde{g}^2)$. Expanding $(k_+,k_-,k,k_0)$ in \eqref{A:k} up to $O(\tau^2)$ with the corresponding coefficients $(c_0, c_1,c_2,d_0,d)$,
\begin{align}\label{C:k_exp}\nonumber
k_+&=c_0+c_1\tau+c_2\tau^2\\\nonumber
    &=4\tilde{g}\sin\phi+[2\tilde{g}\cos\phi]\tau-[\tilde{g}\sin\phi]\tau^2+O(\tau^3)\\\nonumber
    k_-&=c_1\tau+c_2\tau^2\\\nonumber
    &=[2\tilde{g}\cos\phi]\tau-[\tilde{g}\sin\phi]\tau^2+O(\tau^3)\\
    k&=d\tau=-2\tilde{g}^2\tilde{\Delta}\tau\\
    k_0&=d_0\tau=2\tilde{\Delta}\tau\nonumber
\end{align}
and the square root 
\begin{align}\label{C:sqrt}
    \sqrt{\frac{1}{1-ik\eta}}=\sqrt{\frac{1+ik\eta}{1+k^2\eta^2}}\approx1+\frac{1}{2}ik\eta.
\end{align}
With those series, the exponent of $I[k,k_+,k_0]$ in \eqref{B:I(k)} is expanded
\begin{align}\label{C:exponent}\nonumber
    &\frac{ik}{1-ik\eta}\left(q^2+\frac{k_+}{k}q+i\frac{k^2_+\eta}{4k}\right)+ik_0\\\nonumber
    &=-\frac{1}{4}\eta c_0^2+iqc_0\\\nonumber
    &+i\left(-\frac{1}{2}\eta c_0c_1+qc_1+q^2d+d_0\right)\tau\\
    &+\left(-\frac{1}{4}\eta c^2_1-\frac{1}{2}\eta c_0c_2+iqc_2\right)\tau^2+O(\tilde{g}^3)
\end{align}
Similarly, all the other $I[\cdot,\cdot,\cdot]$ can be expanded by replacing $(k,k_+,k_0)$ with the relevant parameters. The Holevo quantity $\chi(\rho)$ in \eqref{Holevo single} is determined by $\mu$ and $E(\beta)$, so what we need is to compute $\mu(\tau)$ in \eqref{mui}. By substituting \eqref{C:k_exp}, \eqref{C:sqrt} and \eqref{C:exponent} into \eqref{B:I(k)} and \eqref{B:I(dk)}, $\mu(\tau)$ is expanded up to $O(\tau^2)$ and $O(\tilde{g}^2)$,
\begin{align}\label{C:mu_squared_expansion}\nonumber
   &\mu^2(\tau)\\
   &= 1-\frac{1}{2}\tau^2\eta c^2_1-\frac{1}{8}\tau^2\eta d_0^2c^2_0\cos^2(qc_0/2)\\\nonumber
   &+O(\tau^3)+O(\tilde{g}^3)\\
   &=1-2\tau^2\eta^2\tilde{g}^2[\cos^2\phi+4\tilde{\Delta}^2\sin^2\phi\cos^2(2q\tilde{g}\sin\phi)]\nonumber
\end{align}
and hence
\begin{align}\label{C:mu_expansion}\nonumber
    \mu(\tau)&=\sqrt{1-2\tau^2\eta^2\tilde{g}^2[\cos^2\phi+4\tilde{\Delta}^2\sin^2\phi\cos^2(2q\tilde{g}\sin\phi)]}\\\nonumber
    &=1-\tau^2\eta^2\tilde{g}^2[\cos^2\phi+4\tilde{\Delta}^2\sin^2\phi\cos^2(2q\tilde{g}\sin\phi)]\\
    &+O(\tau^4).
\end{align}
Keeping in mind $\mu(0)=1$, $\mu'(0)=0$,
\begin{align}\label{C:chi_derivatives}\nonumber
    \chi(\mu=1)&=0\\
    \left.\frac{\partial\chi}{\partial\mu}\right|_{\mu=1}&=0\\
    \left.\frac{\partial^2\chi}{\partial\mu ^2}\right|_{\mu=1}&=\frac{E^2(\beta)}{2}\{\log_2[1-E(\beta)]-\log_2[1+E(\beta)]\}\nonumber
\end{align}
\begin{align}\label{C:chi_O(2)}\nonumber
    \chi(\rho)&\approx\frac{E^2(\beta)}{4}\{\log_2[1-E(\beta)]-\log_2[1+E(\beta)]\}\mu''(0)\\
    &\approx\frac{E^2(\beta)}{2}\log_2\frac{1+E(\beta)}{1-E(\beta)}\\\nonumber
&\times\tau^2\eta^2\tilde{g}^2[\cos^2\phi+4\tilde{\Delta}^2\sin^2\phi\cos^2(2q\tilde{g}\sin\phi)]\\\nonumber
&\approx\frac{E^2(\beta)}{2}\log_2\frac{1+E(\beta)}{1-E(\beta)}(\cos^2\phi+4\tilde{\Delta}^2\sin^2\phi)\eta^2\tilde{g}^2\tau^2.\nonumber
\end{align}

\end{document}